\documentclass[aip,jap,amsmath,amssymb,preprint]{revtex4-1}

\usepackage{graphicx}
\usepackage{dcolumn}
\usepackage{bm}
\usepackage[mathlines]{lineno}

\begin{document}


\title[]{The inter-sublevel optical properties of a spherical quantum dot-quantum well with and without a donor impurity}

\author{Hatice Ta\c{s}}
\email{h.tas@live.com}
\affiliation{Department of Physics, Faculty of Science, Sel\c{c}uk University, Campus 42075 Konya, Turkey}

\author{Mehmet Sahin}
\email{mehmet.sahin@agu.edu.tr}
\affiliation{Department of Physics, Faculty of Science, Sel\c{c}uk University, Campus 42075 Konya, Turkey}
\affiliation{Department of Material Science and Nanotechnology Engineering, Abdullah G\"{u}l University, Kayseri, Turkey}


\begin{abstract}
In this study, we have investigated the inter-sublevel optical properties of a core/shell/well/shell spherical quantum dot with the form of quantum dot-quantum well heterostructure. In order to determine the energy eigenvalues and corresponding wave functions, the Schr\"{o}dinger equation has been solved full numerically by using shooting method in the effective mass approximation for a finite confining potential. The inter-sublevel optical absorption and the oscillator strength between ground ($1s$) and excited ($1p$) states have been examined based on the computed energies and wave functions. Also, the effect of a hydrogenic donor impurity, located at the center of the multi-shell spherical quantum dot (MSQD), has been researched for different core radii ($R_1$), shell thicknesses ($T_s$) and well widths ($T_w$) in certain potential. It is observed that the oscillator strengths and the absorption coefficients are strongly depend on the core radii and layer thicknesses of the MSQD.
%
\end{abstract}

\maketitle

\section{Introduction}

During the last years there has been a tremendous research area on low dimensional semiconductor systems such as quantum wells, quantum wires and quantum dots (QDs).\cite{lin,per} Low dimensional semiconductor systems display various physical properties, different from higher dimensional systems for instance, the discrete energy structure, the absorption spectrum (both intra- and inter-band) are expected here to be a series of discrete lines.\cite{Mil}

Recent improvements in the nanostructure technology have made possible to prepare zero dimensional semiconductor nanostructures, called quantum dots, by using different methods such as molecular-beam epitaxy (MBE).\cite{Kas} As it is well known, when dimensions of the structure are diminished, the quantum mechanical effects become more important and hence these effects improve the performance of many electronic and optoelectronic devices. The attracting features of QDs are that their many properties (energy levels, wave functions, density of states, etc.) can be controlled by the shape, size and composition of the structure. These features are similar to the properties of real atoms\cite{Hol} and therefore QDs are named as artificial atoms. A great number of studies on quantum dot systems such as GaAs/AlGaAs, InGaAs/GaAs\cite{Mir}, InAs/InP\cite{Ust,Pett}, and InAs/In(Ga,As)\cite{Gru} have been reported both experimentally and theoretically for a wide range of dot sizes and shapes. The studies on QDs open a new field in both basic physics and chemistry, and present a wide range of potential applications for optoelectronic devices such as optical and electro-optic modulators\cite{Ree,Qua}, inter-band lasers \cite{Kri}, inter-subband long wavelength detectors.\cite{Jian}

A hydrogenic impurity problem in a QD is a very useful model in understanding of many electronic and optical properties of these kind of structures. A donor impurity atom has a one more electron than that required to make a chemical bond with neighboring atoms in a semiconductor materials. If a donor impurity has an additional one electron, such impurity is known as hydrogenic one because of that it is very similar to a hydrogen atom.\cite{har} The presence of an impurity in a QD changes its effective potential and this affects both energy spectrum and optical transitions because of the Coulomb interaction between the electron and the impurity and is very important in low dimensional semiconductor physics.\cite{Sad} Also, some device properties, such as transition energy, can be tuned by the impurity. Therefore, the effects of impurities on the optical and electronic properties of semiconductor heterostructures draw attention most of the authors and so a number of theoretical and experimental studies have been reported.\cite{Zhu,Por,Bos,Bose,Tkac}

As is well known, when an electron is stimulated by a photon, the electron makes a vertical transition from an initial level to a final one. If the levels are in the same band such as conduction band, this kind of transition is called as inter-subband or intra-band transition. On the other hand, if the transitions occur in between valance and conduction bands, these transitions are described as inter-band ones. Similar kinds of transitions take place in QDs too. Today, the most of optoelectronic devices fabricated from QDs have been built on such transitions. Hence, investigation of the optical absorption properties of QDs are very important in condensed matter and applied physics.\cite{Lees,Klim,Mac,Sau} Li and Xia \cite{Li}, and Li et al. \cite{Lis} have calculated both inter-band and intra-band optical transitions for InAs/GaAs coupled quantum dots. Han et al.\cite{Han} have calculated the inter-subband optical transition of QDs in a quantum well in the framework of the effective mass envelope function theory. Yilmaz and Safak\cite{Yil}, have examined the oscillator strengths of inter-subband transitions for the electron in CdS/SiO2 QD with an on-center donor impurity. Bassani and Buczko \cite{Buc} have showed that how to determine the resonant transition and their lifetimes and how to calculate the oscillator strengths for inter-sublevel optical transitions from the ground state. De Souso et al.\cite{Des} have studied the influence of the Stark effect on the intra-band transitions due to its potential application in infrared detection and emission. The band structure and inter-sublevel hole transitions in valance band of SiGe/Si QDs have been investigated theoretically by Lin and Singh \cite{Linyy}, using eight-band \textbf{k}$\cdot$\textbf{p} model. Gondar and Comas \cite{Gon} have obtained the selection rules for the optical transitions and they reported the oscillator strengths for dipole-allowed transitions in a QD with semi-spherical geometry. Jiang et al.\cite{Jia} have performed theoretical studies on inter-sublevel transitions in two QD systems.

The recent developments in fabrication technology make possible the production of multi-shell quantum heterostructures.\cite{Akt} The experimental studies carried out on these structures are mostly fulfilled based on inter-band exciton transitions.\cite{maz} However, although there are some theoretical studies reported on MSQDs in the literature, the studies are related to the electronic properties of these structures such as energy levels and impurity binding energies.\cite{Boz} Recently, Sahin et al.\cite{sah3} has showed that the electronic shell structure of MSQD is changing and reordering with shell thicknesses. They have demonstrated that the reordering in the electron-conduction band shell structure can be completely different from the hole-valance band shell structures. In our previous work\cite{tas}, we have performed a detailed investigation of the electronic properties of a core/shell/well/shell multi-layered spherical quantum dot for both ground and excited states. To the best of our knowledge, the optical properties of MSQD have not been investigated yet. The controllability of the layer(s) thicknesses may provide different advantages in the production of new generation opto-electronic devices working based on inter-sublevel optical transitions.

The main goal of this study is to research the optical properties such as inter-sublevel absorption and the oscillator strength between ground and excited states of a MSQD structure. These properties have been studied for cases with and without an on-center hydrogenic donor impurity. The calculations have been done for various core radii, barrier thicknesses and well widths for same potential confinement. All results have been presented and discussed comparatively for both cases with ($Z=1$) and without the impurity ($Z=0$).

The rest of the paper is organized as follow: In the next section, we introduce the considered model and its theory. In section \ref{sec:res_dis}, the results of the calculations and their probable physical reasons are discussed. In the last section, a brief conclusion is presented.

\section{Model and Calculations\label{sec:modcal}}

\begin{figure}
\includegraphics[width=\columnwidth]{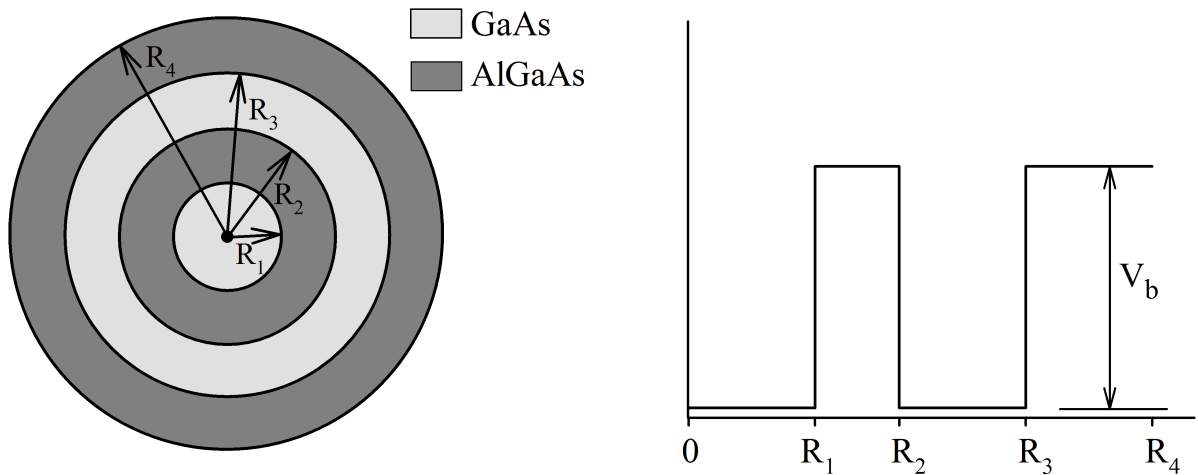}
\caption{\label{fig:1}Schematic representation of a core/shell/well/shell QD and its potential profile.}
\end{figure}

In our study, we deal with a system which includes an electron confined in a MSQD. The considering system consists of two GaAs quantum dot one within the other and each surrounded by AlGaAs layers. The first AlGaAs layer generates a finite potential and it allows the electron to tunnel between the core and well regions. The other AlGaAs layer isolates the whole system from outside. The schematic representation of MSQD and its potential profile are plotted in Fig. \ref{fig:1}. In the considering structure, the core radius is $R_1$, the shell thickness is $T_s=R_2-R_1$ and the well width is $T_w= R_3-R_2$. In the effective mass approximation for a spherically symmetric quantum dot, the single particle Schr\"{o}dinger equation is written as

\begin{widetext}
\begin{equation}
\label{eq1}
\left[ { - \frac{{\hbar ^2 }}{2}\vec \nabla _r \left({\frac{1}{{m^* \left( r \right)}}\vec \nabla _r } \right) - \frac{{Ze^2 }}{\kappa(r) r} + \frac{{\ell \left( {\ell  + 1} \right)\hbar^2 }}{{2m^* \left( r \right)r^2 }}+ V(r)} \right]R_{n,\ell } \left( r \right) = E _{n,\ell} R_{n,\ell } \left( r \right).
\end{equation}
\end{widetext}

\noindent
Here, $\hbar$ is reduced Planck constant, $m^{\ast}(r)$ is the position-dependent electron's effective mass, $Z$ is the charge of the impurity, $\kappa(r)$ is the position dependent dielectric constant, $\ell$ is the angular momentum quantum number, $V(r)$ is the finite confining potential, $E_{n,\ell}$ is the electron energy eigenvalue for specified principle quantum number ($n$) and angular momentum quantum number ($\ell$), $R_{n,\ell}(r)$ is the radial wave function of the electron. It should be noted that the $Z=0$ and $Z=1$ correspond to a case without and with a hydrogenic donor impurity, respectively. The mathematical expression of the confining potential is

\begin{equation}
 \label{eq2}
 V(r)=\left\{{\begin{array}{l}
 0,\,\,\,\,\,\,\,\,\,r \leq R_1 \,\,\, $and$ \,\,\, R_2\leq r \leq R_3\\\\
 V_b,\,\,\,\,\,\,R_1<r<R_2 \,\,\, $and$ \,\,\, r>R_3 \\
 \end{array}} \right. .
\end{equation}

In order to determine the single particle energy levels and corresponding wave functions, Eq.(\ref{eq1}) is solved full numerically by the shooting method. As is well known, this technique converts an eigenvalue problem to an initial-value problem. For this purpose, Hamiltonian operator is discretized on a uniform radial mesh in 1D using the finite differences, then Eq.(\ref{eq1}) can be reduced to an initial-value equation by means of

\begin{widetext}
\begin{equation}
\label{eq3}
R_{n,\ell}(i+1)=\left (\frac{r}{r+h}\right) \left[2+\frac{2m^*h^2}{\hbar^2} \left(V(i) + \frac{\ell(\ell+1) \hbar^2}{2m^*r^2}-E _{n,\ell} - \frac{Ze^2}{\kappa(r) r}\right)\right]R_{n,\ell}(i) - \left(\frac{r-h}{r+h} \right)R_{n,\ell}(i-1),
\end{equation}
\end{widetext}

\noindent
where $i$ is the index of mesh points, $h$ is the distance between two mesh points and it is chosen as 0.005. The details of this method can be found in Ref. \onlinecite{har}.

The photon absorption process can be described as an optical transition that takes place from an initial state to a final one with assisted by a photon. The optical absorption calculations for the inter-sublevel transitions are based on Fermi's golden rule derived from time dependent perturbation theory. The inter-sublevel optical absorption coefficient is given as\cite{Sah}

\begin{equation}
\label{eq4}
\alpha(\hbar\omega)=\frac{16\pi^2\beta_{FS}N_{if}}{n_r V_{con}} \hbar \omega |z_{fi}|^2 \delta (E_f-E_i- \hbar \omega),
\end{equation}

\noindent
where $n_r$ is the refractive index of the semiconductor and this value is 3,15 for GaAs , $V_{con}$ is volume of the confinement regions, $\hbar\omega$ is the incident photon energy, $\beta_{FS}$ is the fine structure constant and its value is 1/137, $N_{if}=N_i-N_f$ is the difference of the number of electron between the initial and final state, and $E_i$ and $E_f$ are energy eigenvalues of the initial and final state, respectively. $z_{if}$ is the dipole matrix element between the initial and final states and the $\delta$ is broadening parameter.

In spherical quantum heterostructures, the selection rule, $\Delta\ell=\pm1$, determines the final state of the electron after the absorption. Hence, the $1s$ state is taken as the initial state and the $1p$ level is considered as the final state. The total wave function of these levels are determined from multiplication of the radial wave functions with the spherical harmonics (i.e. $R_{nl}(r)Y_{\ell m} (\theta,\phi)$).

The dipole matrix element for transitions between $1s-1p$ levels is

\begin{equation}
\label{eq5}
\left| {z_{fi}} \right|^2 = \frac{1}{3}\left|{\int\limits_0^\infty {R_{f}(r) r^3 R_{i}(r) dr} } \right|^2.
\end{equation}
$R_{i}(r)$ and $R_{f}(r)$ are radial wave functions of the initial and final states, respectively. The factor $\frac{1}{3}$ comes from the integration of the spherical harmonics. In addition, the $\delta$ function in absorption equation is replaced by a narrow Lorentzian by means of

\begin{equation}
\label{eq6}
\delta \left( {E_f  - E_i  - \hbar \omega } \right) = \frac{{\hbar \Gamma }} {{\pi \left[ {\left( {\hbar \omega  - \left( {E_f  - E_i } \right)} \right)^2  + \left( {\hbar \Gamma } \right)^2} \right]}},
\end{equation}
where the term of $\hbar\Gamma$ is maximum width in semi height of the Lorentzian and its value is taken 6.4 meV in this study.

Other quantity in the study of optical properties is the determining of oscillator strengths. The oscillator strength gives the information about magnitude of the absorption. That is, the amount of the oscillator strength is directly proportional to the absorption coefficient. For $1s-1p$ transitions the oscillator strength is

\begin{equation}
\label{eq7}
O_{fi}=\frac{2m^*}{3\hbar^2}(E_f-E_i)\left| {\int\limits_0^\infty {R_{f}(r) r^3 R_{i}(r) dr}} \right|^2.
\end{equation}

\section{Results and Discussion \label{sec:res_dis}}

The atomic units have been used throughout the calculations, where Planck constant $\hbar=1$, the electronic charge $e=1$ and the bare electron mass $m_0=1$. Effective Bohr radius is $a_0\simeq100${\AA} and effective Rydberg energy is $R_y\simeq5.25$ meV. The material parameters have been taken as $m_{GaAs}$=0.067$m_0$, $m_{AlGaAs}$=0.088$m_{0}$, $V_b$=228 meV, $\kappa_{GaAs}$=13.18, $\kappa_{AlGaAs}$=12.8. Also the effective masses of electrons inside GaAs and AlGaAs are $m_1^*$ and $m_2^*$, and the dielectric constants are $\kappa_1$ and $\kappa_2$, respectively. The position-dependent effective mass and the dielectric constant may be defined as follows\cite{Buc}

\begin{eqnarray}
\label{eq8}
\nonumber
m^\ast(r) = \left\{ {\begin{array}{l}
 1,\,\,\,\,\,\,\,r \leq R_1 \,\,\, $and$ \,\,\, R_2\leq r \leq R_3 \\\\
 \frac{m_2^\ast }{m_1^\ast },\,\, R_1<r<R_2 \,\,\, $and$ \,\,\, r>R_3 \\
 \end{array}} \right. \\ \nonumber
\\
\kappa(r) = \left\{ {\begin{array}{l}
 1,\,\,\,\,\,r \leq R_1 \,\,\, $and$ \,\,\, R_2\leq r \leq R_3 \\\\
 \frac{\kappa_2 }{\kappa_1 },\,\,R_1<r<R_2 \,\,\, $and$ \,\,\, r>R_3 \\
 \end{array}} \right. .
\end{eqnarray}

In Eq. \ref{eq4}, all terms except $N_{if}$, $\hbar\omega$, $V_{con}$ and Lorentzian are constants and these constants have no effect on the determining of absorption shape. In the resonant absorption case which occurs when the photon energy becomes equal to the energy difference between the levels, the Lorentzian is also a constant, $1/\pi\hbar\Gamma$. As the optical absorption coefficient is directly proportional with the dipole matrix element, it is inversely proportional with the volume of the confinement regions. Hence, the overlapping, volume of the confinement regions and the energy levels become more effective on the absorption coefficient. Therefore, the probable physical reasons in the absorption properties will be discussed as dependent on these parameters. Similarly, the oscillator strength also depends on the energy difference between considered levels and the overlapping of the wave functions of these levels.

It should be noted that the overlapping of the radial wave functions is calculated by means of

\begin{equation}
\label{eq9} \Lambda ={\int\limits_0^\infty  {R_{f}(r) R_{i}(r) r^2 dr} }.
\end{equation}

\subsection{Effect of the core radius on the optical properties}

\begin{figure}
\includegraphics[width=\columnwidth]{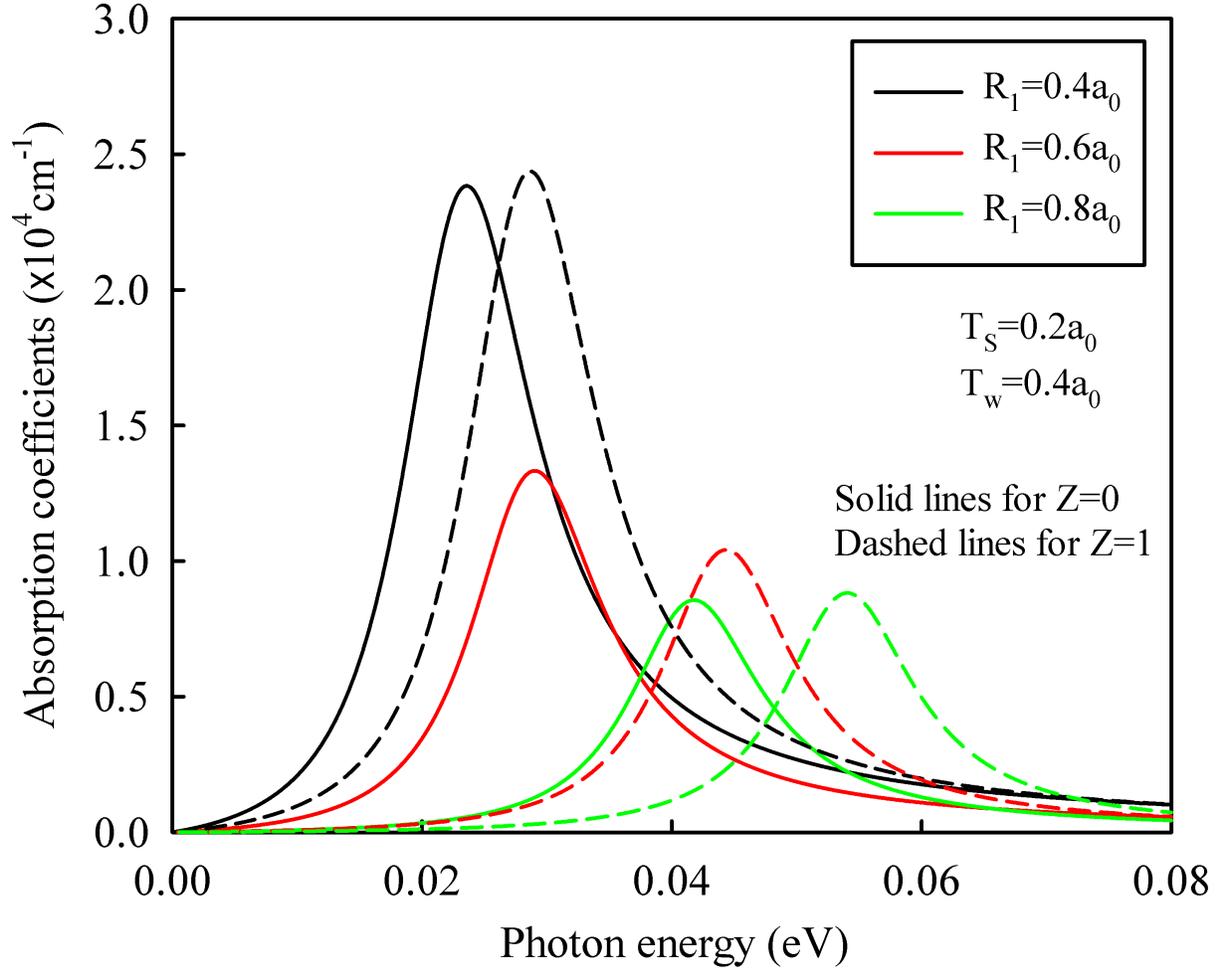}
\caption{\label{fig:2} (Color online) Variation of the absorption spectra as a function of incident photon energy for different core radii in cases of $Z=0$ and $Z=1$. The shell thickness is $T_s=0.2\ a_0$, the well width is $T_w=0.4\ a_0$.}
\end{figure}

\begin{figure}
\includegraphics[width=\columnwidth]{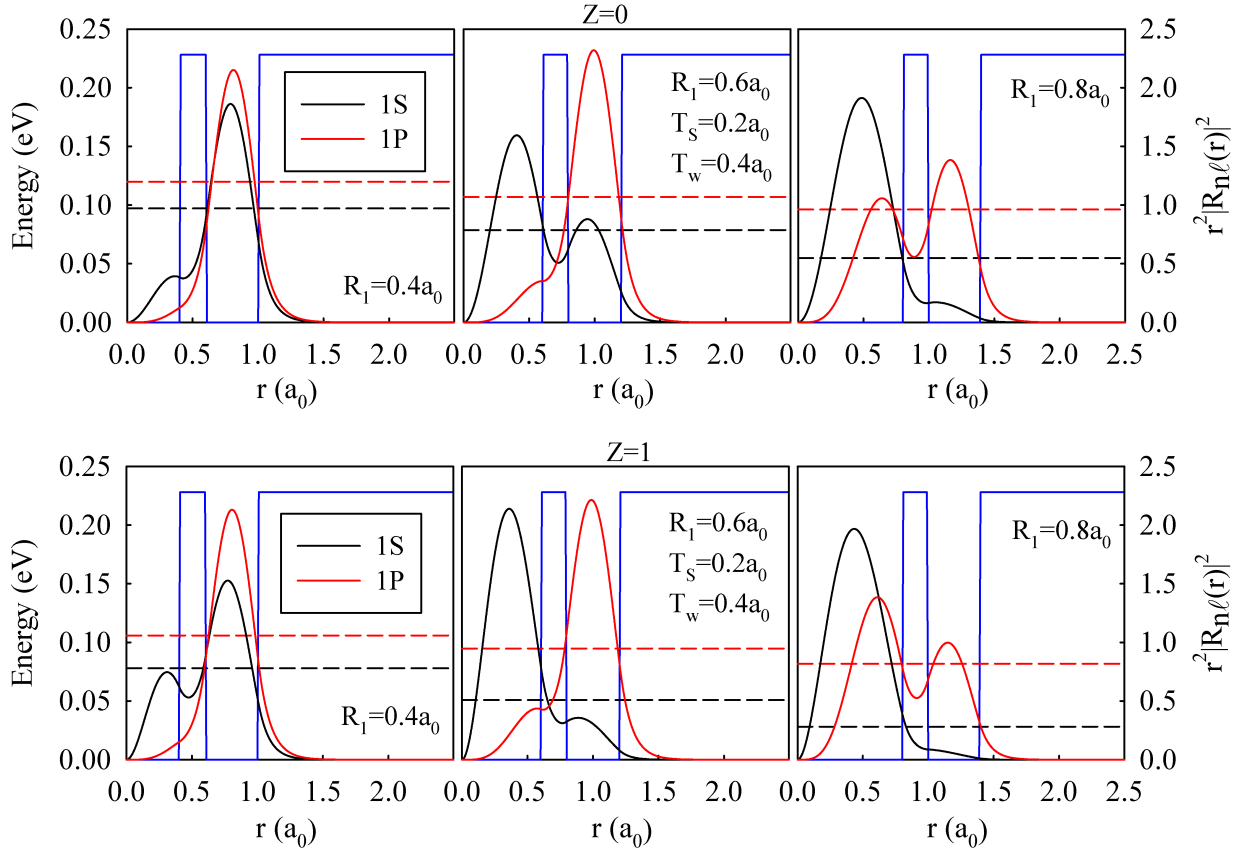}
\caption{\label{fig:3} (Color online) Ground and excited states probability distributions as a function of the $r$ for different core radii. The top panel for $Z=0$ and the bottom panel for $Z=1$ case. The energies of levels are showed with dashed lines.}
\end{figure}

Figure \ref{fig:2} shows the absorption coefficient variation with the incident photon energy for $Z=0$ and $Z=1$ cases for different core radii and constant shell thickness ($T_s$) and well width ($T_w$). As it is known, when an electron is stimulated by a photon, the electron absorbs the photon and moves up in the vertical direction to higher energy levels. If the incident light energy equals to the energy difference between the levels, this energy is called as resonance transition energy which corresponds to the peak energy of the absorption coefficient. In Fig. \ref{fig:2}, the peak energies of the absorption coefficient for $Z=1$ case have higher values (blue shift) than those of $Z=0$ case at all core radii. This situation results from the attractive Coulomb potential of the donor impurity. Because, in case of $Z=1$, the impurity pulls ground ($1s$) level down more than excited ($1p$) level and hence the energy difference between $1s$ and $1p$ levels becomes larger comparing with $Z=0$ case as seen evidently from energy levels of $1s$ and $1p$ states in Fig.\ref{fig:3}. On the other hand, although the absorption coefficient values are almost same for $Z=0$ and $Z=1$ cases at $R_1=0.4$ and $0.8\ a_0$, it is larger for $Z=0$ than in case of $Z=1$ for $R_1=0.6\ a_0$. This can be explained by means of the overlapping of the wave functions given in Fig. \ref{fig:3} for $Z=0$ and $Z=1$ cases. That is, while the overlapping of the wave functions are almost same in both $Z=0$ and $Z=1$ cases for $R_1=0.4$ and $0.8\ a_0$ as seen from right and left panels of Fig. \ref{fig:3}, the overlapping is stronger in case of $Z=0$ than that for $Z=1$ case because the finding probability of ground and excited levels are confined in different region in latter case by comparison with $Z=0$ case as seen from middle panel of Fig. \ref{fig:3}.

\begin{figure}
\includegraphics[width=\columnwidth]{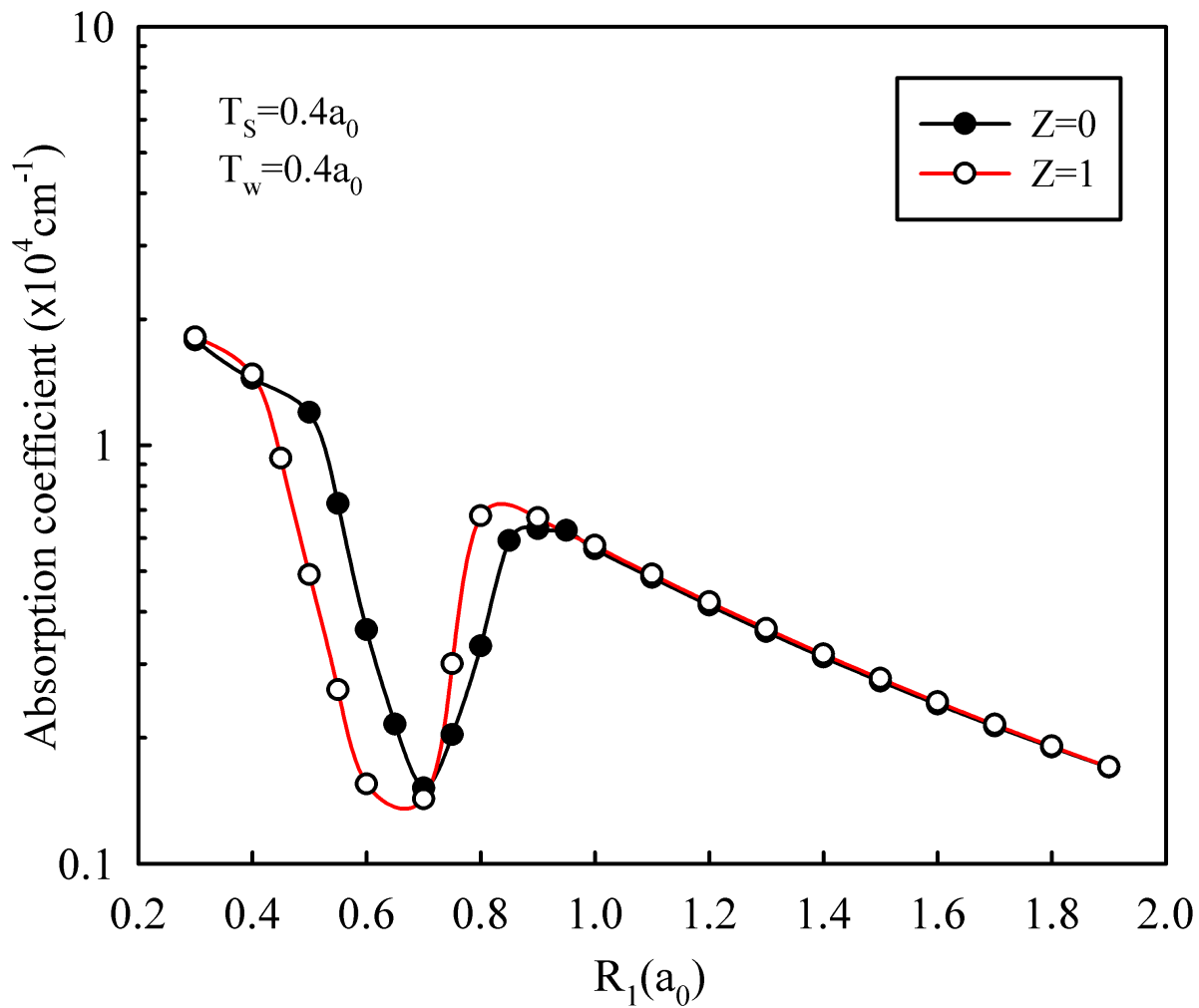}
\caption{\label{fig:4} (Color online) Absorption coefficient peak values as depending on core radius for $Z=0$ and $Z=1$ cases. The shell thickness is $T_s=0.4\ a_0$, and well width is $T_w=0.4\ a_0$.}
\end{figure}

The variation of peak values of the absorption coefficient with the core radius may provide some important information in device applications. For this purpose, the absorption peak values are plotted as a function of $R_1$ for $Z=0$ and $Z=1$ cases in Fig. \ref{fig:4}. The structure parameters are given on the figure. According to the figure, the absorption coefficient presents the downward tendency in general for both cases except that sudden decreasing and increasing. This sudden decreasing and increasing in the absorption coefficient with $R_1$ can be explained as depending on overlapping of the wave functions. When the $R_1=0.3\ a_0$ ground and excited states wave functions are localized in the well region. Therefore, the overlapping is strong and hence the dipole matrix element has a big value. In this situation, the optical transition takes place in the well region. With the increase of the core radius, the overlapping decreases and so the absorption coefficient decreases too. In this case, while the wave function of ground state is localized in the core region, the excited states wave function is confined in the well one. In large core radii ($R_1>0.9\ a_0$), the wave functions of ground and excited states are localized in the core region completely and therefore the absorption coefficient becomes large for the cases of $Z=0$ and $Z=1$. The absorption takes place in the core region anymore. On the other hand, although the overlapping increases, the decrease in the absorption coefficient can be explained by the increase of the confinement region volume which results from the increase of $R_1$.

\begin{figure}
\includegraphics[width=\columnwidth]{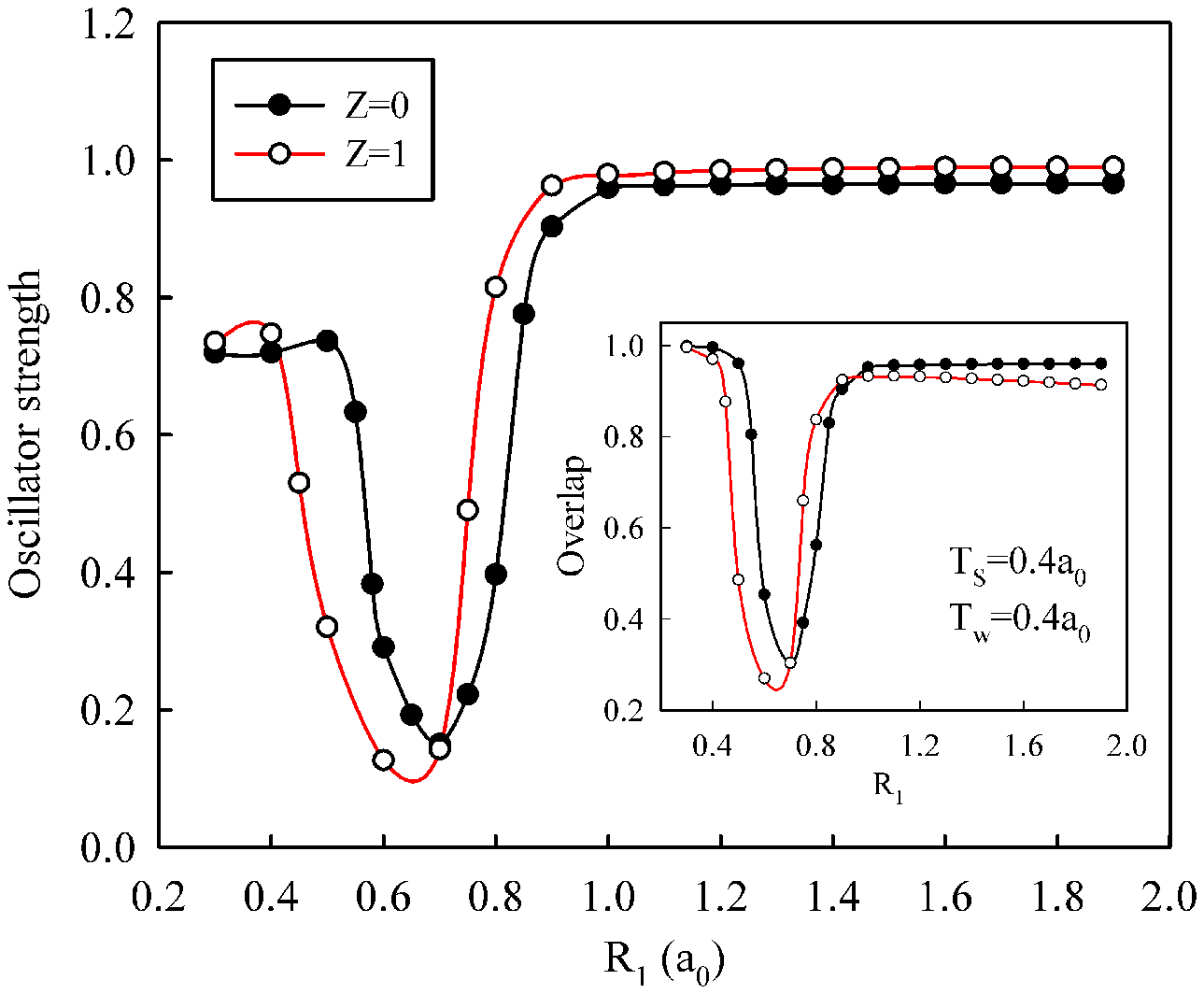}
\caption{\label{fig:5} (Color online) The variation of oscillator strength as a function of core radius for $Z=0$ and $Z=1$ cases. The structure parameters are same with the previous figure. The inset figure shows the overlap of the wave functions.}
\end{figure}

The oscillator strength variation of the transition as a function of the core radius is given in Fig. \ref{fig:5} for $Z=0$ and $Z=1$ cases. The inset figure shows the overlaps of the wave functions calculated from Eq. \ref{eq9}. From the figure, we see that the oscillator strength does not show a considerable change in small core radii. After that, it exhibits a sudden fall and rise tendency similar as the absorption coefficient. It is remains fixed about 1.0 after $R_1\geq1a_0$ with further increasing of $R_1$. The oscillator strength is strongly dependent on both energy difference between the levels and overlapping of the wave functions as it is mentioned before. Although the energy difference is little in small values of $R_1$, overlapping of the wave functions is strong as seen from the inset and thus the oscillator strength is also higher. The overlapping exhibits a decrease tendency rapidly with increasing $R_1$. This decreasing situation is faster in $Z=1$ case. In contrast, when $R_1>0.6\ a_0$, the increase in overlap of $Z=1$ case becomes more quickly than that of $Z=0$. In addition to this, the energy difference gets some more in $Z=1$ case as seen from Fig. \ref{fig:4}. In parallel with these, the oscillator strengths of the impurity are larger after certain $R_1$ value.

\subsection{Effect of the shell thickness on the optical properties}

\begin{figure}
\includegraphics[width=\columnwidth]{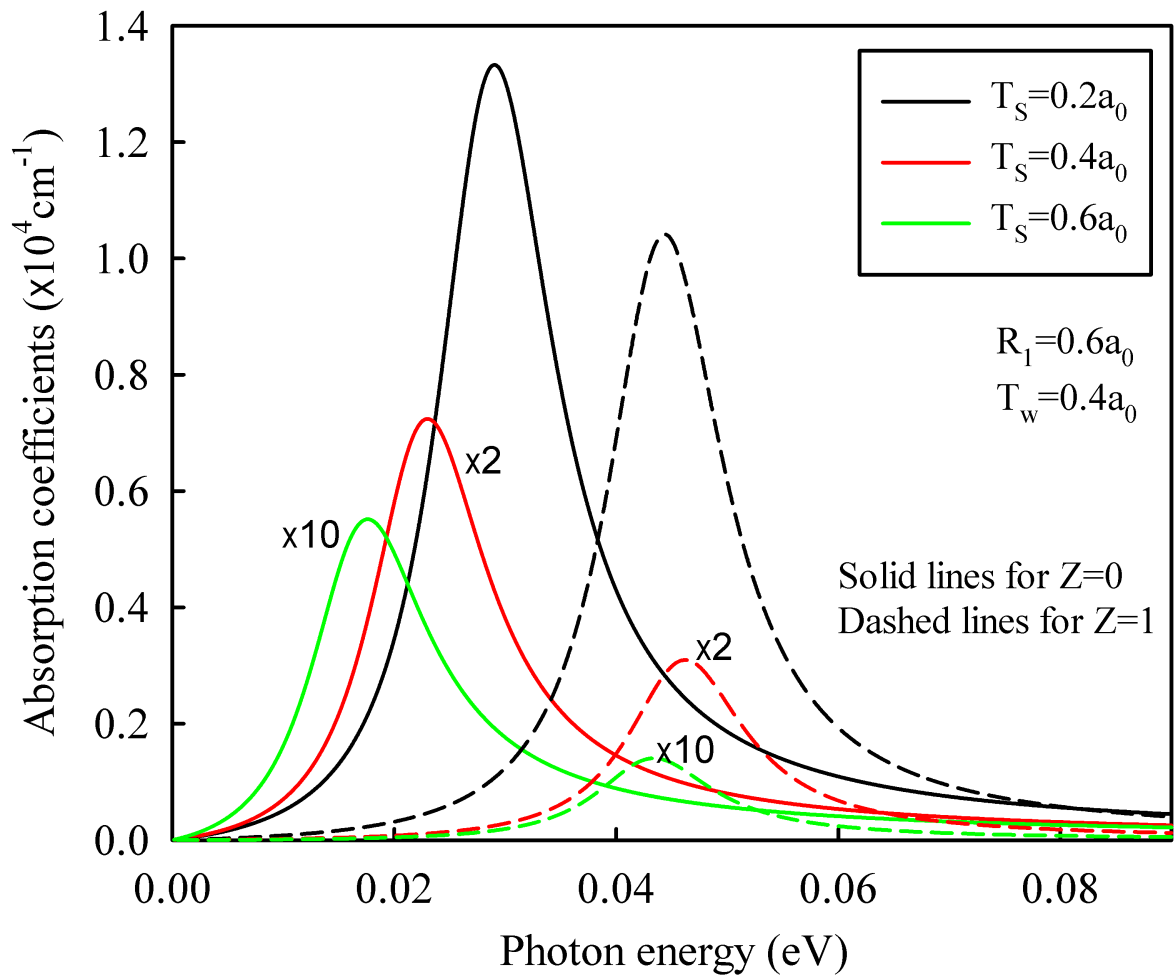}
\caption{\label{fig:6} (Color online) Absorption coefficient as a function of photon energy at different shell thicknesses for $Z=0$ and $Z=1$ cases. Other structure parameters are specified on the figure.}
\end{figure}

\begin{figure}
\includegraphics[width=\columnwidth]{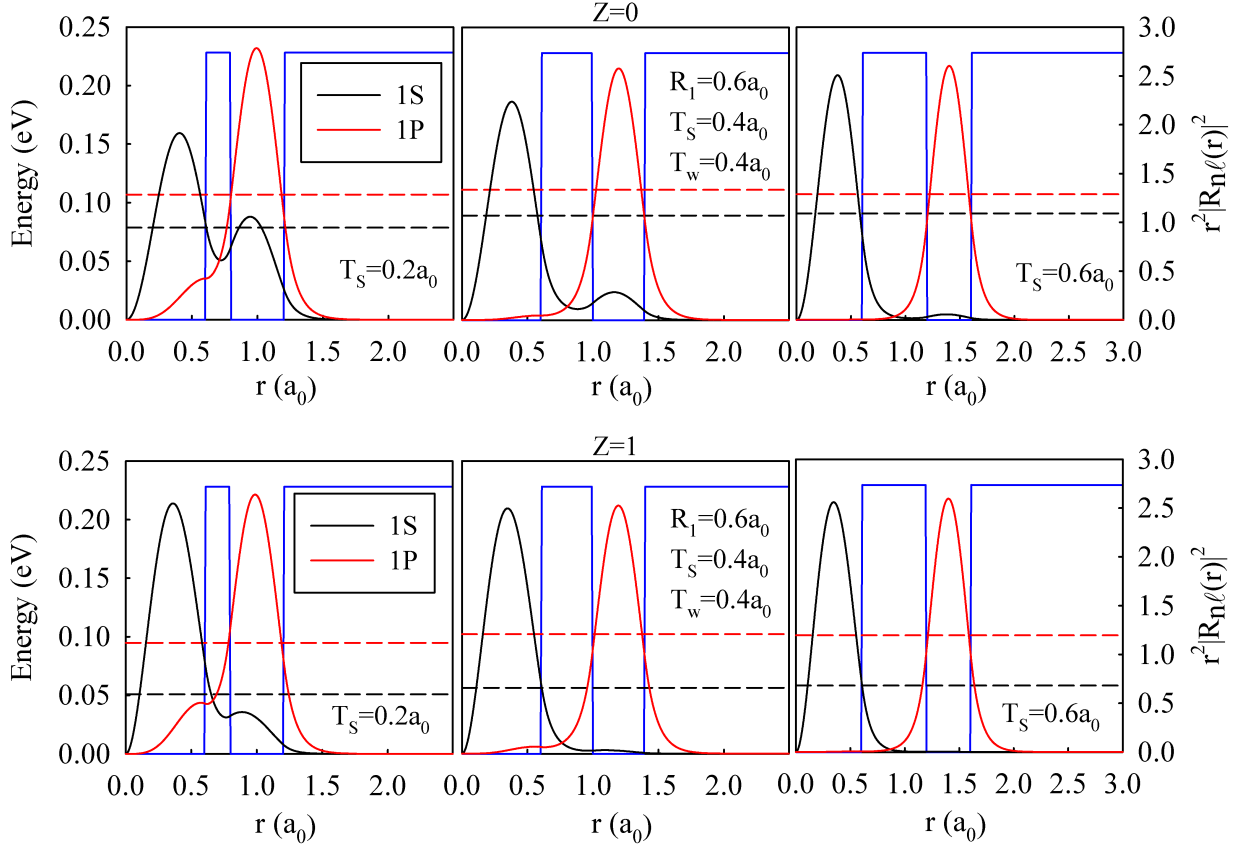}
\caption{\label{fig:7} (Color online) Ground and excited states probability distributions as a function of $r$ for different shell thicknesses in case of $Z=0$ (top panel) and for $Z=1$ (bottom panel).}
\end{figure}

Fig. \ref{fig:6} shows the variation of absorption coefficient spectra as depending on the photon energy for different shell thicknesses and the other parameters of considering structure are $R_1=0.6\ a_0$ and $T_w=0.4\ a_0$. The figure is comparatively plotted for $Z=0$ and $Z=1$ cases. As seen from the figure, the absorption coefficients of $Z=0$ case are higher than that of $Z=1$ case for all shell thicknesses. When the shell thickness increases, the absorption coefficient decreases very rapidly for both $Z=0$ and $Z=1$ cases. Therefore, the absorption coefficients are multiplied by 2 and 10 for $T_s=0.4$ and $T_s=0.6$ $a_0$, respectively. In addition, although there are no significant changes in the absorption peak energies (i.e. transition energy) for $Z=1$ case, these changes are notable and downward tendency with increasing shell thickness in case of $Z=0$. The probability distributions of dealing structure for $Z=0$ and $Z=1$, which are top and bottom panel of Fig \ref{fig:7}, respectively, are given for $T_s=0.2$, 0.4, and 0.6 $a_0$ for certain core radius and well width. As seen from the figure, ground and excited states wave functions become localized in different region with the increase of the $T_s$. For $T_s=0.2 \ a_0$, while a part of ground state finding probability is in the well region in case of $Z=0$, because of attractive Coulomb potential of the impurity it is almost completely in core region for $Z=1$ case. On the other hand, the finding probability of excited states are localized in the well region for both $Z=0$ and $Z=1$ cases. This is because the probability of tunneling, between core and well regions, decreases with the increase of the shell thickness. This process makes weak the overlapping of the wave functions of ground and excited states. Also, when we look at the top panel of Fig. \ref{fig:7}, we observe that the energy difference between the $1s$ and $1p$ states decreases with the increase of shell thickness and hence, the peak energies of the absorption coefficient decreases (red shift) as seen in Fig. \ref{fig:6}. When we compare the absence of impurity with the existence of one, in $Z=1$ case, the impurity pulls the $1s$ energy states down, thus the energy difference between ground and excited levels is larger than that of $Z=0$ case. On the other hand, the energy levels remain almost fix with increasing of the shell thickness for $Z=1$ case and hence the absorption peak energies become steady.

\begin{figure}
\includegraphics[width=\columnwidth]{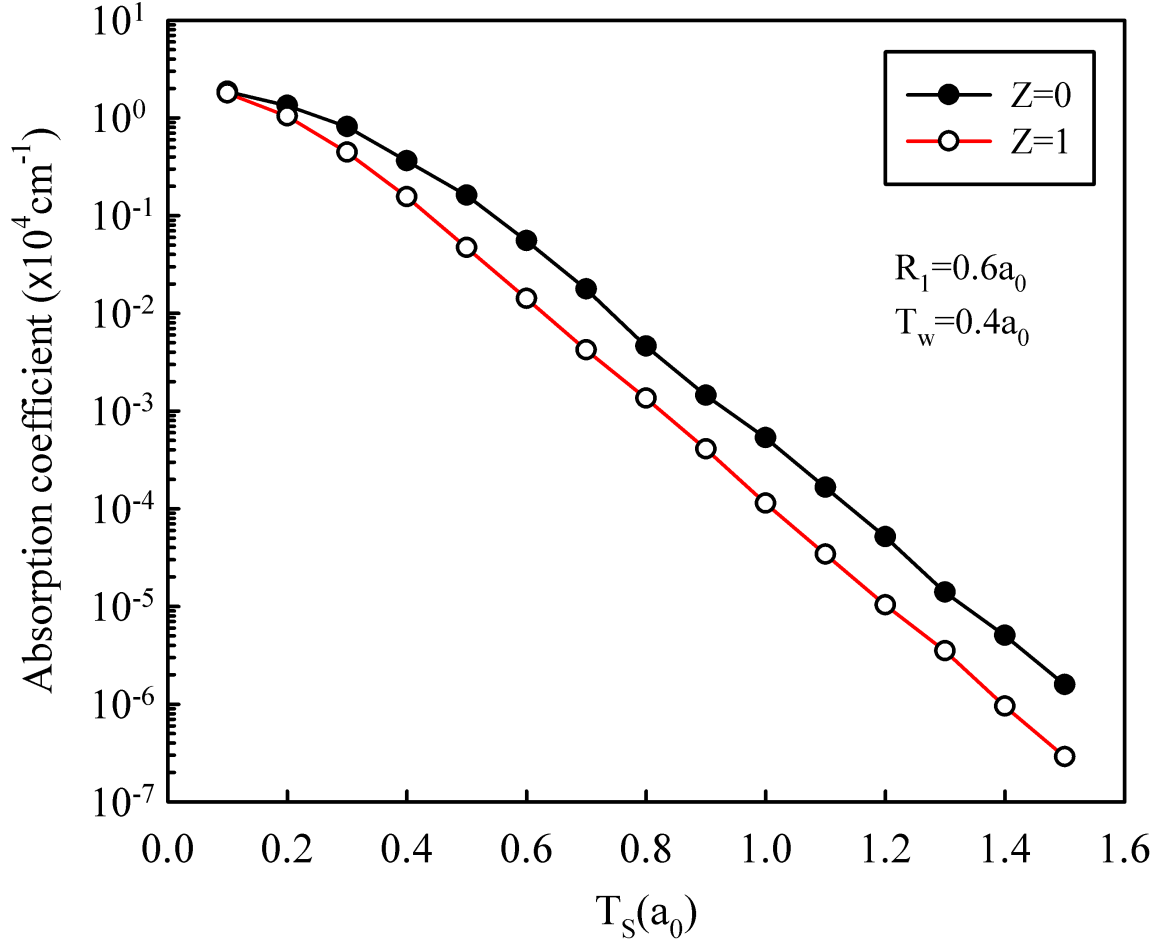}
\caption{\label{fig:8} (Color online) The variation of peak values of the absorption coefficients as depending on $T_s$ for $Z=0$ and $Z=1$. The core radius is $R_1=0.6\ a_0$, and the well width is $T_w=0.4\ a_0$.}
\end{figure}

Fig. \ref{fig:8} shows the variation of the maximum values of the absorption coefficient with the shell thickness, for $Z=0$ and $Z=1$ cases. The parameters of the structure are given on the figure. As is expected, the peak values of absorption coefficient exhibit a downward tendency with the increase of $T_s$. Although there is no change in the confinement region volume, the decrease of the overlapping lead to this downward tendency. Whereas, in small shell thicknesses, the difference of absorption coefficient for with and without the impurity is very small, this difference rises gradually and remains almost constant in large shell thicknesses. Moreover, the values of absorption coefficient for $Z=1$ are lower than that of $Z=0$. In this case, the decrease of the overlapping becomes more effective than the increase of the energy difference between the states for $Z=1$.

\begin{figure}
\includegraphics[width=\columnwidth]{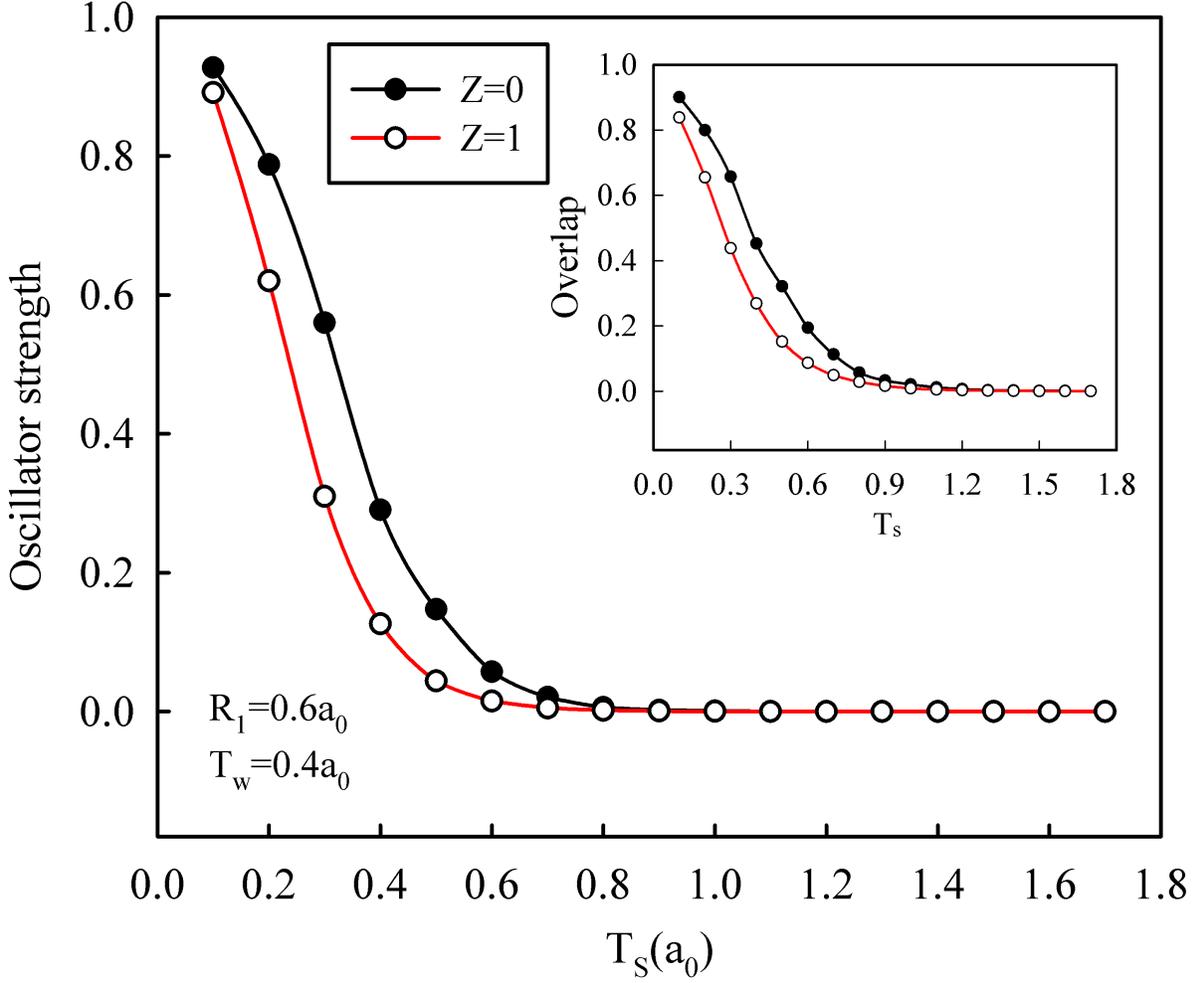}
\caption{\label{fig:9} (Color online) Oscillator strength variation as depending on $T_s$ for $Z=0$ and $Z=1$. The core radius is $R_1=0.6\ a_0$, and the well width is $T_w=0.4\ a_0$. The inset shows the overlap of the wave functions.}
\end{figure}

The oscillator strength of the structure as a function of $T_s$ is given in Fig. \ref{fig:9} for the same parameters with the previous figure. The inset figure shows the variation of the overlap integral with the shell thickness. As seen from Fig. \ref{fig:9}, the oscillator strength has a maximum value at the beginning and reduces rapidly with the increasing of $T_s$, and it goes towards zero with further increasing of the $T_w$ for both $Z=0$ and $Z=1$ cases. When we look at the figure, it is observed that the oscillator strength for $Z=1$ smaller than that for $Z=0$ until a certain value. This case stems from the impurity atom. Because, the impurity pulls the ground state down and so, it localized in the core region while the excited state localized in the well one. Thus, as seen from the inset, the overlapping of the wave functions and as depending on this, the oscillator strength decreases. Also, the increase in $T_s$ causes the decrease in tunneling between the core and well regions. Therefore, these regions do not feel each other.

\subsection{Effect of the well width on the optical properties}

\begin{figure}
\includegraphics[width=\columnwidth]{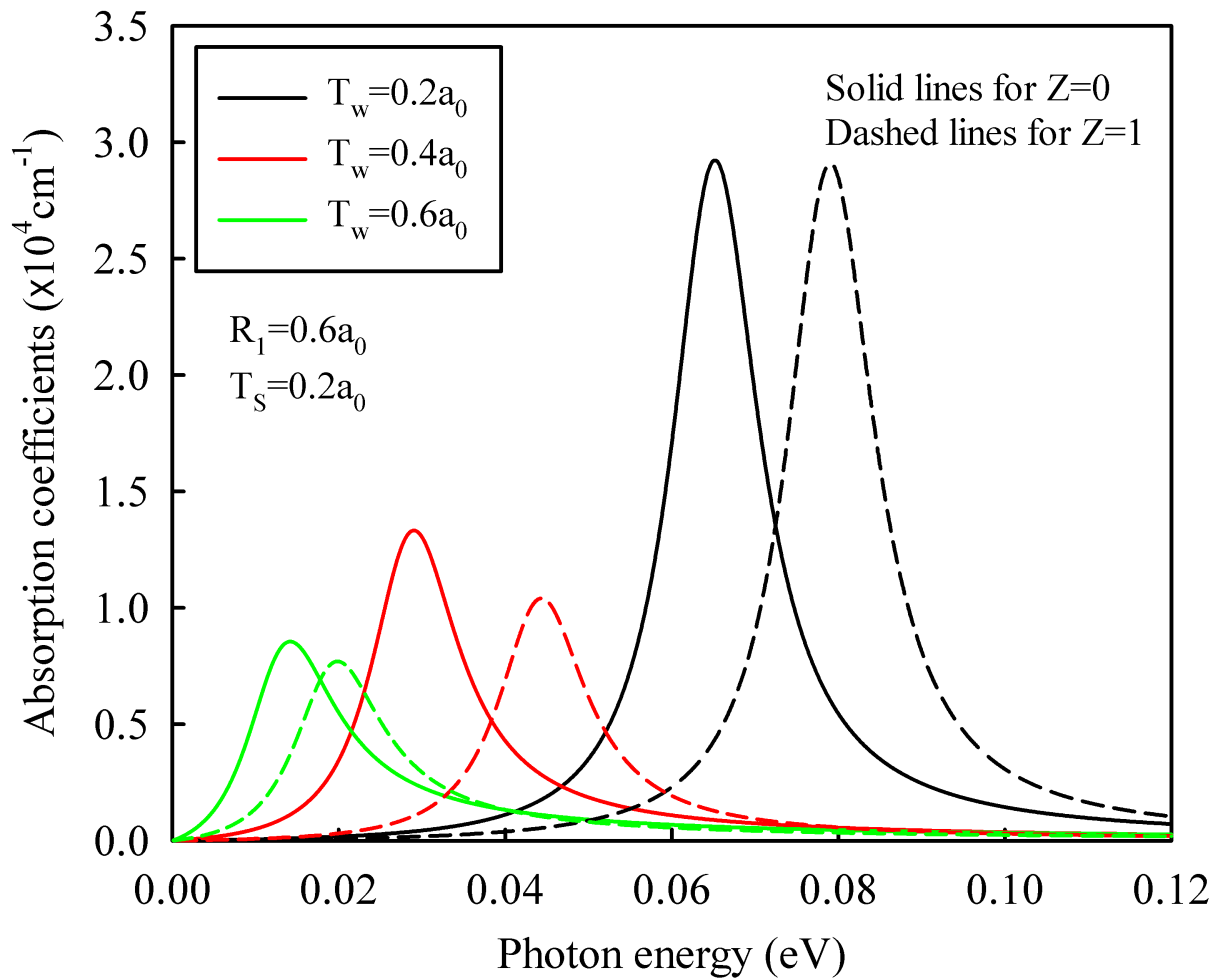}
\caption{\label{fig:10} (Color online) The variation of absorption coefficient as a function of the photon energy for different well widths in cases of $Z=0$ and $Z=1$. The core radius is $R_1=0.6\ a_0$, and the shell thickness is $T_s=0.2\ a_0$.}
\end{figure}

\begin{figure}
\includegraphics[width=\columnwidth]{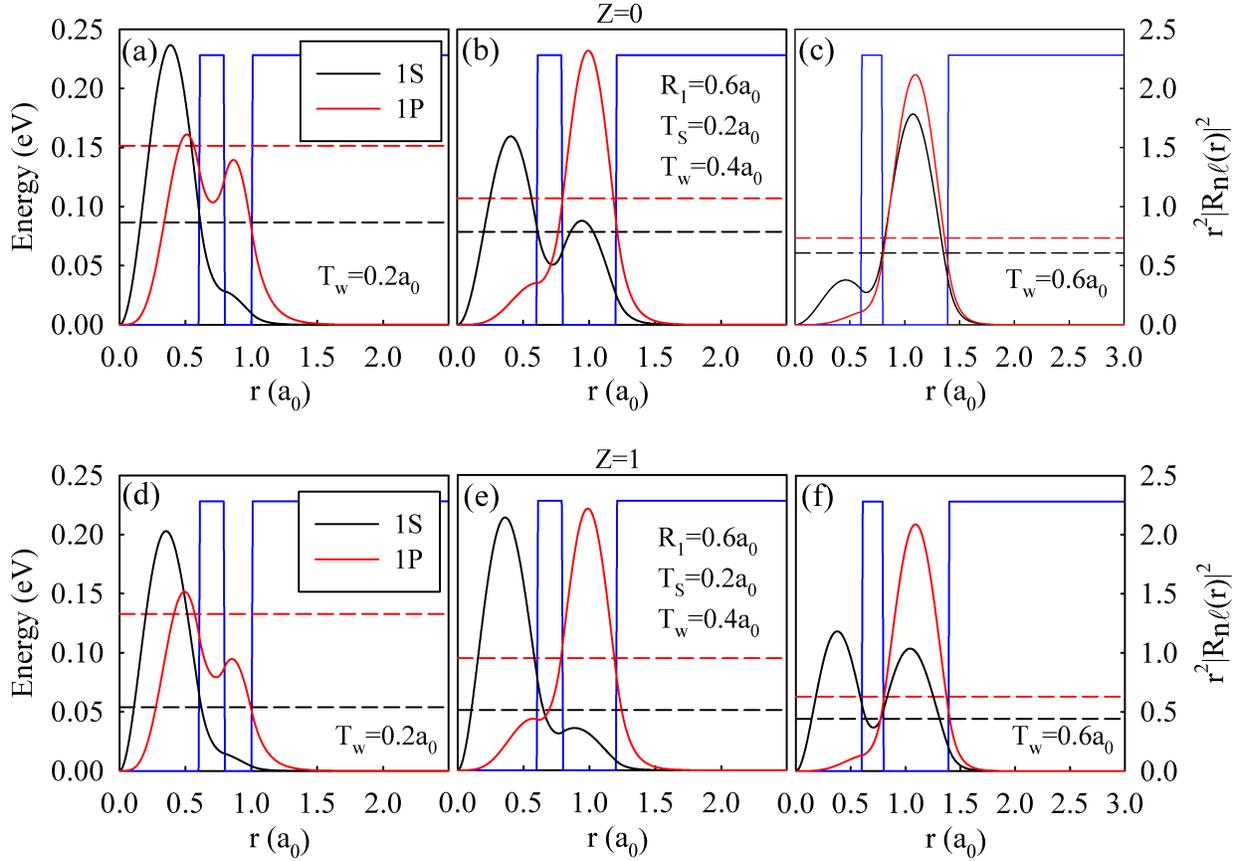}
\caption{\label{fig:11} (Color online) Ground and excited states probability distributions as a function of $r$ for various well widths in cases of $Z=0$ (top panel) and $Z=1$ (bottom panel). The other structure parameters are specified on the figures.}
\end{figure}

The variations of optical absorption coefficient with the well widths may be important for different device applications of MSQDs. For this reason, we have calculated the absorption coefficients depending on the well widths. Figure \ref{fig:10} shows the absorption coefficients as a function of incident photon energy in case of $Z=0$ and $Z=1$ for three different well widths, i.e. $T_w=0.2$, 0.4 and 0.6 $a_0$. The core radius and the shell thickness are chosen as $R_1=0.6\ a_0$ and $T_s=0.2\ a_0$, respectively. When we look at the figure, we observe that although the absorption coefficients of $Z=0$ case are a bit larger than that of $Z=1$, their peak energies of $Z=1$ case are greater (blue shift) than that of $Z=0$ for given well widths. Also, it can be seen that the absorption coefficients and their peak energies are higher for lower $T_w$ and reduce with increase of the $T_w$ for both $Z=0$ and $Z=1$. This can be explained as follows with the assistance of Fig. \ref{fig:11}: since the well width is narrow, the probability densities of the levels are higher in the core region and hence, the overlapping of the wave functions increases relatively with respect to larger well widths (i.e. $T_w\geq0.4\ a_0$) and it leads to the absorption occurs in the core region as seen from Fig. \ref{fig:11} (a) and (d). In this case, the energy difference between the states is also bigger. In addition, the volume of the confinement region is small due to the narrow well width. As a results of these, when $T_w=0.2\ a_0$, both the absorption coefficients and their peak energies become larger. When $T_w=0.4\ a_0$, the absorption coefficients and their peak energies reduce. This is because the energy level of excited state decreases considerably owing to the tunneling of excited state wave functions to the well region for both $Z=0$ and $Z=1$ cases as seen in Fig. \ref{fig:11} (b) and (e). In case of $Z=1$, because of attractive Coulomb potential of the impurity, ground state wave functions almost completely confines in the core region as excited state wave function confines in the well one. This results in small overlapping of the wave functions and so small absorption coefficient as compare with case of $Z=0$. When $T_w=0.6\ a_0$, the probability density of excited states are confined completely in the well region in both $Z=0$ and $Z=1$ cases as seen from Fig. \ref{fig:11} (c) and (f). On the other hand, ground state density of $Z=0$ is localized almost completely in the well region as that of $Z=1$ case is shared between the core and well regions. When $T_w>0.6\ a_0$, the absorption takes place in the well region.

\begin{figure}
\includegraphics[width=\columnwidth]{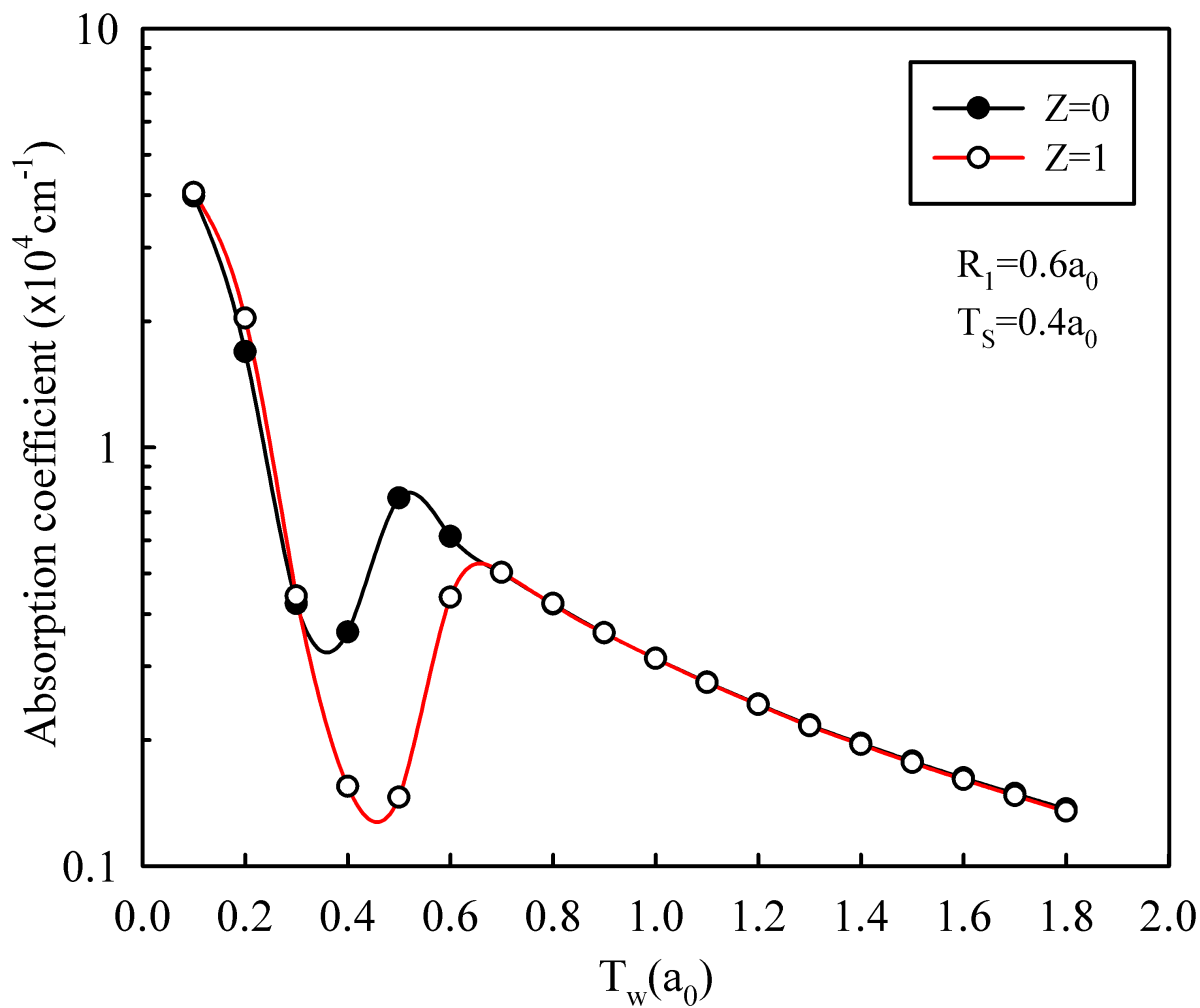}
\caption{\label{fig:12} (Color online) The variation of absorption coefficient peak values as depending on $T_w$ for $Z=0$ and $Z=1$. The core radius is $R_1=0.6\ a_0$ and the shell thickness is $T_s=0.4\ a_0$.}
\end{figure}

\begin{figure}
\includegraphics[width=\columnwidth]{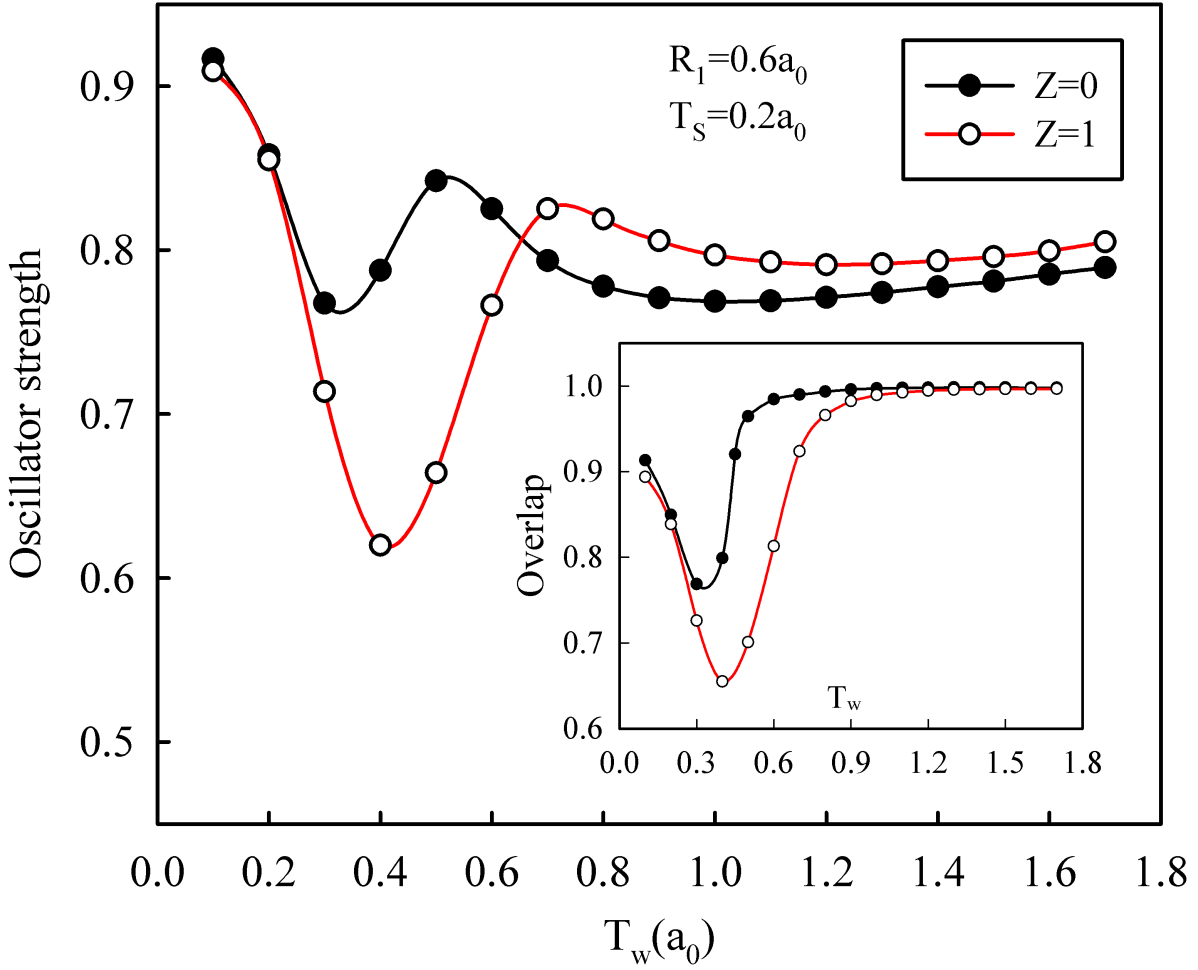}
\caption{\label{fig:13} (Color online) The oscillator strength variation as depending on $T_w$ for $Z=0$ and $Z=1$ cases. The core radius $R_1=0.6\ a_0$, and the shell thickness is $T_s=0.2\ a_0$. The inset shows the overlap of the wave functions.}
\end{figure}

Fig. \ref{fig:12} shows the variation of absorption coefficient peak values as a function of the $T_w$ in cases of $Z=0$ and $Z=1$ for $R_1=0.6\ a_0$ and $T_s=0.4\ a_0$. As expected, the absorption coefficient exhibits a reduce tendency with the increase of the $T_w$ in general. When $T_w$ is about $0.4\ a_0$ for $Z=0$, the absorption coefficient becomes minimum and it increases again with further increasing of $T_w$. In $Z=1$ case, the absorption coefficient reaches its minimum value when $T_w=0.5\ a_0$. These decreases can be explained by means of that the overlapping of the wave functions is weak as seen from the inset given in Fig. \ref{fig:13}. In Fig. \ref{fig:12}, it is observed that the absorption peak values in between $T_w=0.3$ and $0.7\ a_0$ are smaller for $Z=1$ case by comparison with $Z=0$ case. Because, the attractive Coulomb potential of the impurity reduces the overlapping of the wave functions. When $T_w=0.7\ a_0$, the absorption coefficients of both cases have same values and exhibit same treatment. That is, both the core region and the impurity lose of their own effects on the considered quantum states anymore and the optical transitions take place in the well region.

In Fig. \ref{fig:13}, for the structure with the $R_1=0.6\ a_0$ and $T_s=0.2\ a_0$, the variation of the oscillator strengths are given for cases with and without the impurity depending on $T_w$. It is seen that, in the beginning, the oscillator strengths have maximum values and then, they exhibit a decrease tendency with increasing of the $T_w$ and reach minimum values. In the latter case, ground state wave function is localized in the core region, while excited state wave function is confined in the well one. Afterwards, they increase with further increasing of the $T_w$ and decrease again smoothly. As similar to the previous figure, when almost $T_w<0.7\ a_0$, the oscillator strength in the presence of the impurity has smaller value due to the negative effect of the impurity coulomb potential on the overlapping. After this certain value of $T_w$, the oscillator strength is to become higher for $Z=1$ than that of $Z=0$ case, although the overlapping is equally stronger in both cases as observed from the inset of Fig. \ref{fig:13}. The reason of this alteration is because the difference between the energy levels is larger in case of $Z=1$ owing to attractive coulomb potential of the impurity on ground state, especially. We conclude from here, the energy difference becomes more predominant on the oscillator strength than the overlapping.

\section{Conclusion}

In this study, we have calculated the absorption coefficient and oscillator strength for the inter-sublevel transitions in a MSQD heterostructure. Here, $1s$ and $1p$ levels are taken into consideration as ground and first excited states, respectively. The results have been evaluated depending on core radii ($R_1$), shell thicknesses ($T_s$), and well widths ($T_w$) and we have discussed the probable physical reasons behind the results. We have observed that the optical properties are drastically dependent on the layer thicknesses of the structure. We have concluded that many of the optical properties can be controlled by tuning the layer thicknesses for different device applications. To our knowledge, this study is first investigation of the detailed optical properties of a MSQD. We believe that this study will be rather useful and will contribute to the understanding of the optical properties of MSQDs. We hope that this study will also stimulate both experimental and theoretical investigations of the optical properties of MSQDs.

\section*{Acknowledgement}

This study is a part of M.Sc. Thesis prepared by H. Ta\c{s} at Physics Department of Selcuk University.

\end{document}